\documentclass[letterpaper,twocolumn,10pt]{article}
\ifdefined\pdfsuppressptexinfo
\fi
\usepackage{usenix-2020-09}

\usepackage{amsmath}
\usepackage{amssymb}
\usepackage{graphicx}
\usepackage{multirow}
\usepackage{xspace}
\usepackage{subfigure}
\usepackage{booktabs}
\usepackage{array}
\usepackage{makecell}
\usepackage{caption}
\usepackage{float}
\usepackage{algorithm}
\usepackage{algorithmic}
\usepackage{flafter}

\newcommand{\paraspace}{\vspace{0pt}}
\newcommand{\parab}[1]{\paraspace\noindent{\bf #1} }

\newcommand{\needlines}[1]{%
  \par\begingroup
  \dimen0=\pagegoal
  \advance\dimen0-\pagetotal
  \dimen2=#1\baselineskip
  \ifdim\dimen0<\dimen2\newpage\fi
  \endgroup}

\newcommand{\mydesign}{\textsf{VarioPath}}

\begin{document}

\date{}
\title{\mydesign{}: Workload-Aware All-to-All Communication for PCIe GPU Clusters}
\author{%
  \normalfont
  Yao Fei$^{1}$ \quad Jin Fang$^{1}$ \quad Size Zheng$^{2}$ \quad
  Gongming Zhao$^{1}$ \\
  \normalfont Hongli Xu$^{1}$ \quad Jiacheng Zhu$^{1}$ \quad Zhijing Xin$^{1}$ \\[0.35em]
  \normalfont {\small $^{1}$University of Science and Technology of China \qquad
  $^{2}$Tsinghua University} \\[0.2em]
  \normalfont {\small\texttt{\{yao\_fei,fangjin98,zhu\_jc,zhijingxin\}@mail.ustc.edu.cn}} \\
  \normalfont {\small\texttt{\{gmzhao,xuhongli\}@ustc.edu.cn} \qquad
  \texttt{zhengsz@mail.tsinghua.edu.cn}}%
}
\maketitle

\begin{abstract}
AlltoAllv communication is a critical primitive in distributed large-model
inference, particularly for mixture-of-experts (MoE) models. The growing adoption
of PCIe GPU systems for cost-efficient inference makes AlltoAllv performance
on these systems increasingly important. Without a dedicated scale-up
interconnect (e.g., NVLink or Infinity Fabric), PCIe GPU systems carry both
intra-node and inter-node traffic through the PCIe hierarchy, where concurrent
transfers can contend for PCIe link bandwidth. This link contention,
compounded by skewed traffic distributions and dynamic traffic demand,
makes efficient AlltoAllv scheduling challenging.
Existing approaches are either poorly suited to PCIe GPU
systems or incur substantial schedule synthesis overhead that reduces their practicality
in real-world deployments.

We present \mydesign{}, an efficient AlltoAllv scheduling framework for PCIe GPU systems.
It combines an offline topology-aware analyzer with an online demand-aware
scheduler. The analyzer records contention-free transfer patterns as
\emph{AlltoAllv channels} and exploits topology symmetry to build a compact catalog
for efficient search. The online scheduler decomposes each AlltoAllv invocation's
demand across a sequence of channels, adapting to rapidly changing and skewed
traffic while incurring low planning overhead.
Evaluation on four platforms (up to 256 GPUs) shows average AlltoAllv speedups
of $5.88\times$ over FAST and $1.72\times$ over DeepEP. End-to-end
experiments show that \mydesign{} reduces Qwen3 inference latency by up to
27.2\% and Wan2.1 generation latency by 6.1\%.

\end{abstract}

\section{Introduction}
AlltoAllv is a critical communication primitive for distributed model inference.
In expert parallelism (EP)~\cite{rajbhandari2022deepspeed,fastalltoall,deepep,shazeer2017moe,lepikhin2021gshard,fedus2022switch}, it dispatches activations to experts and returns their outputs.
In sequence parallelism (SP)~\cite{fang2024usp,gu2024loongtrain}, it redistributes attention tensors across the sequence and head dimensions.
An AlltoAllv invocation may require each GPU to send different volumes of data to different peers.
For example, input-dependent expert routing can produce a skewed traffic distribution~\cite{fastalltoall,hwang2023tutel,he2022fastermoe}, with a few destinations receiving substantially more data than others.
AlltoAllv's many-to-many exchange pattern complicates the coordination of concurrent transfers and limits efficient utilization of interconnect bandwidth~\cite{du2026ecoserve,hwang2023tutel,kim2024tccl,li2019evaluating}.
Consequently, AlltoAllv communication often becomes a major bottleneck in distributed large-model inference.

This communication bottleneck is even more pronounced on PCIe GPU systems.
In tightly integrated GPU servers (e.g., NVIDIA HGX H100~\cite{nvidiafabricmanager}), a dedicated NVLink/NVSwitch scale-up fabric~\cite{NVLink} provides high-bandwidth, non-blocking intra-node GPU connectivity~\cite{slechta2024nvlinkinference}.
By contrast, PCIe systems connect GPUs and NICs through shared PCIe switches and links.
These interconnects vary across server designs and do not guarantee non-blocking communication~\cite{li2019evaluating,shah2023taccl,kim2024tccl}.
Their bandwidth is also substantially lower than that of dedicated scale-up networks.
Despite these limitations, PCIe systems are widely deployed for inference because their availability and customizability help control infrastructure cost and meet different resource requirements~\cite{du2026ecoserve,kim2024tccl}.
For example, Alibaba and ByteDance deploy PCIe-connected NVIDIA T4 and L40S GPUs in production clusters comprising thousands of GPUs~\cite{weng2022mlaas,zhu2025megascale}.
Cloudflare also deploys multiple types of PCIe GPUs for its Workers AI service~\cite{cloudflare2024gen12}.

On these widely deployed PCIe GPU systems, AlltoAllv presents three challenges:
\emph{link contention}, \emph{traffic skew}, and \emph{dynamic demand}.
First, the typically non-Clos interconnect can concentrate traffic on the same
bottleneck links~\cite{li2019evaluating,shah2023taccl}, while the absence of a
separate scale-up network makes intra-node GPU communication and NIC-based
inter-node traffic compete for the same PCIe resources.
Traffic therefore concentrates on some links, creating hotspots while leaving
others underutilized and reducing aggregate throughput.
Second, AlltoAllv demand can be highly skewed, requiring some GPUs to
transfer much more data than others. This imbalance keeps certain GPUs and
PCIe links busy after others have finished, creating a communication tail
that delays the entire invocation even without link contention.
Finally, dynamic AlltoAllv demand makes static schedules impractical and requires
schedulers to adapt in real time.

Prior work~\cite{fastalltoall,deepep,shah2023taccl,liu2024teccl,cao2025syccl,wang2020blink} primarily targets tightly integrated servers with dedicated scale-up networks.
For example, FAST~\cite{fastalltoall} and DeepEP~\cite{deepep} rebalance AlltoAllv traffic by introducing additional intra-node transfers over dedicated, high-bandwidth scale-up networks.
On PCIe GPU systems, however, these approaches exacerbate link contention because the additional intra-node transfers compete with NIC traffic for the same shared PCIe links.
General-purpose libraries such as NCCL~\cite{NCCL} and MSCCL~\cite{MSCCL} incur little runtime scheduling overhead, but their static scheduling strategies cannot effectively adapt to dynamic AlltoAllv demands in real-world workloads.
Only a few systems, including Blink~\cite{wang2020blink}, \mbox{BlueConnect}~\cite{cho2019blueconnect}, and TCCL~\cite{kim2024tccl}, explicitly optimize collective communication for PCIe GPU systems.
However, they focus primarily on regular collectives such as AllReduce~\cite{patarasuk2009bandwidth} and AllGather~\cite{bruck1997alltoall}, often using ring-based algorithms.
Their optimizations therefore do not directly apply to AlltoAllv.
MILP-based schedulers~\cite{shah2023taccl,liu2024teccl,cao2025syccl,wu2025swot} jointly model network topology and communication demand but can take seconds to hours to synthesize a schedule.
Because AlltoAllv demand changes across invocations, repeatedly running this synthesis is impractical for real-world workloads.

Addressing the above challenges while maintaining low scheduling overhead is difficult
on PCIe GPU systems. This is because avoiding link contention requires time-consuming
topology analysis, while dynamic AlltoAllv demand requires schedules to adapt to every invocation.
\emph{Our key insight is that the PCIe topology remains stable
and determines which concurrent transfer patterns avoid link contention.}
These patterns can therefore be identified in advance and recorded as reusable
\emph{AlltoAllv channels}, each comprising concurrent GPU pairs and their communication paths.
These channels distill the topology's communication capabilities into reusable scheduling
primitives. Channels offer different distributions of transfer capacity, making them suitable
for different communication demands. On the demand side, skewed AlltoAllv demand can be
divided into chunks and mapped to a sequence of channels in real time. This decomposition balances traffic
across different paths to mitigate communication tails. Overall, this separation makes effective,
low-overhead AlltoAllv scheduling possible.

Based on this insight, we present \mydesign{}, combining an offline topology-aware analyzer with an online demand-aware scheduler.
To avoid enumerating and storing an exponentially growing channel space,
the offline analyzer exploits topology symmetry to represent channels
compactly as \emph{channel families}.
It then organizes these families in a tree-structured catalog, enabling successive
filtering of candidates.
At runtime, the scheduler maps demand chunks to these families, jointly selecting channels and allocating transfer volumes to their paths.
The scheduler prioritizes communication bottlenecks when matching channels
to the current demand. Reusing the offline catalog allows it to achieve near-optimal schedules
with low scheduling overhead.
We make the following contributions:
\begin{itemize}
    \item We present \mydesign{}, an AlltoAllv scheduling framework
    for PCIe GPU systems. By separating topology-dependent offline analysis
    from demand-dependent online scheduling, \mydesign{} enables topology-aware
    and demand-adaptive AlltoAllv scheduling with low runtime overhead.

    \item We distill topology information into the AlltoAllv channel abstraction,
    with each channel representing a contention-free concurrent transfer pattern.
    The offline analyzer exploits topology symmetry to organize channels into a
    compact tree-structured catalog for efficient search, while the bottleneck-aware online scheduler decomposes each
    invocation's demand and allocates it across a sequence of channels.

    \item We validate \mydesign{} on four PCIe GPU platforms, from single-node
    servers to 256 GPUs, using collective benchmarks and end-to-end Qwen3 and
    Wan2.1 inference. \mydesign{} consistently
    outperforms state-of-the-art systems, achieving up to a $5.88\times$ AlltoAllv
    speedup and reducing inference latency by up to 27.2\%.
\end{itemize}

\section{Background and Motivation}
\label{sec:background}
\begin{figure*}[!t]
  \centering
  \setlength{\subfigcapskip}{-3pt}
  \subfigure[Fastest-path assignment]{%
    \includegraphics[width=.323\textwidth,trim=0 4.25pt 0 0,clip]
      {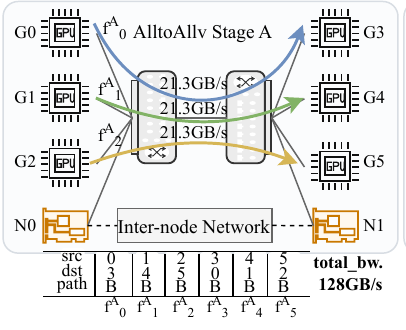}%
    \label{fig:pcie-stage-a}}
  \hfill
  \subfigure[Path-only adaptation]{%
    \includegraphics[width=.323\textwidth,trim=0 4.25pt 0 0,clip]
      {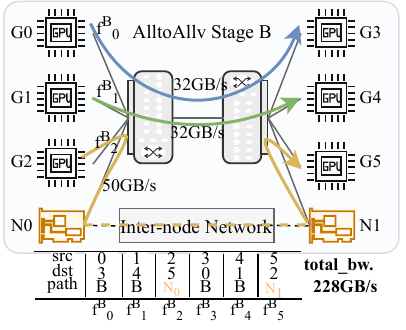}%
    \label{fig:pcie-stage-b}}
  \hfill
  \subfigure[Joint peer--path selection]{%
    \includegraphics[width=.323\textwidth,trim=0 4.25pt 0 0,clip]
      {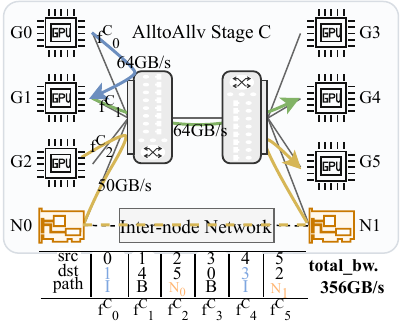}%
    \label{fig:pcie-stage-c}}
  \caption{\textbf{Joint peer--path selection avoids PCIe contention.}
  Panels (a)--(c) compare fastest-path assignment, path-only adaptation, and
  joint peer--path selection, whose modeled aggregate bandwidths are 128, 228,
  and 356~GB/s, respectively.
  Each panel depicts an endpoint-disjoint stage from the same AlltoAllv
  invocation. Each flow $f$ represents a logical transfer together with its selected
  physical path. Transfers not selected in the current stage remain
  for later stages. For clarity, each topology draws only
  $f_0^X$--$f_2^X$, while its table and aggregate bandwidth include all six
  logical transfers selected in the stage. The table below each topology lists
  every source's destination and selected path.
  $\mathrm{I}$, $\mathrm{B}$, and $\mathrm{N}_0/\mathrm{N}_1$ denote a
  same-switch PCIe path (PIX), an inter-switch PCIe path (PXB), and NIC paths
  through NIC0/NIC1, respectively. For simplicity, we use nominal PCIe and NIC
  bandwidths of 64 and 50~GB/s, respectively.}
  \label{fig:pcie-shared-links}
\end{figure*}

\begin{figure}[!t]
  \centering
  \includegraphics[width=\columnwidth]
    {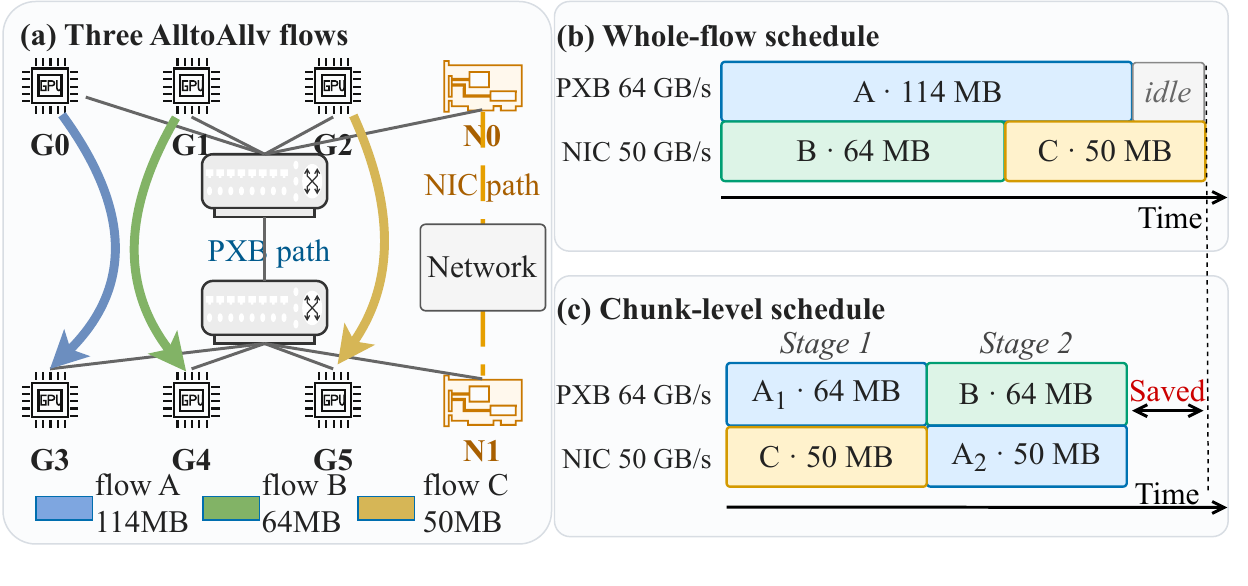}
  \caption{\textbf{Chunk-level path remapping mitigates communication tails.}
  (a) Three cross-switch transfers can use either a 64-GB/s PXB or a 50-GB/s
  NIC path; block width in (b)--(c) denotes modeled transfer time. (b)
  Whole-flow binding creates imbalanced path loads: B and C serialize on the
  NIC, while the PXB path finishes early and becomes idle. (c) Chunking A enables
  path switching in Stage~2, balancing load and eliminating the tail.}
  \label{fig:chunk-path-remapping}
\end{figure}

\subsection{AlltoAllv Scheduling Challenges on PCIe Systems}

An AlltoAllv invocation is described by a matrix of pairwise data volumes,
where each entry specifies the data sent from one source rank to one destination
rank. Support for independently sized rank-pair transfers makes AlltoAllv well
suited to key data-redistribution operations in distributed large-model
inference, including token dispatch and output return in expert parallelism
(EP)~\cite{fastalltoall,deepep,gale2023megablocks} and attention-tensor
redistribution in sequence parallelism
(SP)~\cite{fang2024usp,gu2024loongtrain}. In MoE inference, token-to-expert
assignments are recomputed at every layer, producing a new AlltoAllv demand
matrix. With per-layer execution times often below one
millisecond~\cite{zhao2025deepseekinsights,longcatflash,zhang2025comet}, the
scheduler may need to adapt to each new demand matrix on a similarly short
timescale. Otherwise, scheduling can
delay the critical path and erase the
latency gains of a better communication plan.

Following prior pairwise-exchange schedules~\cite{thakur2005optimization,netterville2022visual,fastalltoall,bruck1997alltoall},
all-to-all exchanges are commonly divided into \emph{endpoint-disjoint} stages.
Within each stage, every rank sends to at most one peer and receives from at most
one peer. Because each rank receives from at most one peer within a stage, this
scheduling structure avoids receiver-side incast, where multiple senders compete
for a receiver's ingress bandwidth and cause severe
congestion~\cite{gangidi2024rdma,lei2025flash}. However, even with
pairwise-exchange scheduling, PCIe GPU systems still present two additional
AlltoAllv scheduling challenges:

\parab{Shared-link contention limits AlltoAllv throughput.}
A non-blocking NVLink/NVSwitch fabric~\cite{li2019evaluating} supports
such endpoint-disjoint logical transfers without internal-link oversubscription.
PCIe-based systems lack an independent scale-up fabric and therefore cannot
guarantee contention-free execution of endpoint-disjoint transfers.
Specifically, this limitation stems from two properties of
PCIe systems. First, intra-node P2P and NIC-based inter-node logical transfers
traverse the same PCIe hierarchy and may contend for GPU-facing links and
other shared PCIe links~\cite{shah2023taccl}.
Second, PCIe fabrics are typically
non-Clos~\cite{clos1953study,li2019evaluating,shah2023taccl,kim2024tccl}, so
logical transfers that are endpoint-disjoint may still share bandwidth-limited
PCIe links. Figure~\ref{fig:pcie-stage-a} shows that endpoint-disjoint logical
transfers can still contend when their paths converge on the same PCIe link.
The resulting
contention reduces the
effective rates of all paths using
that bottleneck and increases the stage completion time while alternative paths
remain underutilized. Accumulated across stages, this loss directly lowers
end-to-end AlltoAllv throughput.

\parab{Skewed demand creates long communication tails.}
AlltoAllv demand may be skewed, with some logical transfers carrying
substantially more data than others. Some paths may consequently carry
disproportionately heavy loads relative to their available rates. These paths
remain active after more lightly loaded paths have finished and become idle, as
illustrated in Figure~\ref{fig:chunk-path-remapping}(b). AlltoAllv completes
only after its last logical transfer, so these stragglers create a long
communication tail that delays the entire invocation.

\subsection{Why Existing Designs Fall Short}

\parab{General-purpose communication libraries.} NCCL~\cite{NCCL}
incurs little runtime scheduling overhead but uses a static AlltoAllv schedule
that does not adapt to changing communication demand.
MSCCLang~\cite{cowan2023mscclang} schedules logical transfers by controlling
their ordering and concurrency, while leaving their physical paths fixed. This
is insufficient for PCIe systems, where logically independent transfers may
still contend on shared physical links. Consequently, neither system can adapt
its schedule to rapidly changing AlltoAllv demand.

\parab{Demand-aware AlltoAllv schedulers for hierarchical systems.} Some
systems optimize AlltoAllv for servers with hierarchical architectures that
provide independent scale-up and scale-out networks. FAST~\cite{fastalltoall}
uses additional intra-node forwarding to balance scale-out traffic, whereas
DeepEP~\cite{deepep} uses it to reduce inter-node traffic.
On PCIe GPU systems without an independent scale-up
fabric, however, these schemes may fail to
deliver their intended gains and instead
exacerbate PCIe contention: their additional intra-node transfers and NIC-based
inter-node traffic traverse the same switched PCIe hierarchy, increasing load on
shared PCIe links. The resulting contention reduces the effective bandwidth of
concurrent transfers and ultimately lowers end-to-end AlltoAllv throughput.

\parab{Topology-aware synthesis.} TACCL~\cite{shah2023taccl},
TE-CCL~\cite{liu2024teccl}, and SCCL~\cite{cai2021synthesizing} synthesize
a schedule by jointly considering a specified communication demand and a
detailed model of physical paths and shared resources. Prior systems report
synthesis times ranging from minutes to
hours~\cite{shah2023taccl,liu2024teccl,cai2021synthesizing}.
Regenerating such a schedule for every dynamic AlltoAllv invocation is therefore
impractical, particularly at scale.

\subsection{Design Observations and Key Insight}

\parab{Observation 1: Avoiding PCIe link contention requires joint peer--path selection.}
Within an AlltoAllv stage, peer matching determines which logical transfers
execute concurrently, while path assignment determines which physical
resources they use. These decisions jointly determine contention: optimizing
either one in isolation can still map multiple transfers onto the same
bottleneck link while leaving alternative paths underutilized.
Figure~\ref{fig:pcie-shared-links} compares three possible stages in an
AlltoAllv schedule:
fastest-path assignment (Stage~A), path-only adaptation (Stage~B), and joint
peer--path selection (Stage~C). In Stages~A and~B, the cyclic
matching~\cite{thakur2005optimization,netterville2022visual,fastalltoall} uses
$k=3$, so rank $i$ sends to $(i+k)\bmod 6$.
Under this peer matching, path-only adaptation raises modeled aggregate
bandwidth from 128 to 228~GB/s but leaves PXB transfers contending.
Stage~C jointly changes matching and paths, combining local PIX,
noncontending PXB, and NIC transfers to eliminate modeled PCIe contention and
reach 356~GB/s.

Inspired by this observation, we use a \emph{channel} to represent the static
peer--path configuration of a stage:
concurrent GPU pairs and one physical path per pair. Intuitively, a
\emph{feasible channel} contains peer--path choices whose transfers can execute
concurrently without causing PCIe link contention, as illustrated by
Figure~\ref{fig:pcie-shared-links}(c). Because feasible channels combine
different peer pairs and path rates, they provide different transfer
capabilities. Their feasibility depends only on the topology, allowing them to
be precomputed offline.

\needlines{4}
\parab{Observation 2: Skewed demand requires adaptive chunk-to-path mapping.}
AlltoAllv demand may be skewed, and each logical transfer may have multiple
candidate paths with different rates. Whole-flow binding can therefore
imbalance path loads even without link contention, leaving faster paths idle
while slower ones determine completion. Figure~\ref{fig:chunk-path-remapping}(b)
illustrates this mismatch: the PXB path becomes idle before the NIC path
finishes. In (c),
splitting transfer~A and remapping its second chunk to NIC balances the two
paths over two rounds. This example shows why path assignments must be
revisited at chunk granularity: between rounds, outstanding traffic can be
remapped to match load to path rates.

\parab{Key insight.}
Together, these observations reveal a natural separation between topology and
demand. The topology changes infrequently and determines which peer--path
configurations can coexist without modeled contention; \mydesign{} precomputes
these configurations as reusable channels. At runtime, \mydesign{} represents
each invocation's demand as a
\mbox{\textbf{\textit{linear combination}}} of channels and assigns
invocation-specific data volumes to the selected paths. This separation
replaces repeated combinatorial topology search with lightweight online demand
decomposition.

\section{Scheduling Model}
\label{sec:scheduling-model}
\parab{Model setup.}
We model an AlltoAllv invocation on a fixed topology comprising a set $V$ of
$N$ GPU ranks and a demand matrix $D\in\mathbb Z_{\ge0}^{N\times N}$, where
$D_{ij}$ denotes the data volume from rank $i$ to rank $j$. Self-traffic is
handled locally and omitted, so $D_{ii}=0$. For each ordered pair $i\ne j$,
let $\mathcal P_{ij}$ denote the nonempty set of available concrete paths.
Each path $p\in\mathcal P_{ij}$ has an estimated rate
$\mathrm{bw}(p)>0$ and uses a set of modeled resources
$\mathcal L(p)\subseteq\mathcal U$. The resource set $\mathcal U$ contains the
directed shared PCIe and NIC-facing resources considered by our model. We
exclude endpoint-local GPU-to-switch links from $\mathcal U$ because the
one-to-one peer matching defined below ensures that no two lanes in a channel
share the same endpoint link in the same direction.

Table~\ref{tab:model-notation} summarizes the notation used in Sections~3
and~4.
\begin{table}[!ht]
  \centering
  \caption{Key notation used in Sections~3 and~4.}
  \label{tab:model-notation}
  \footnotesize
  \setlength{\tabcolsep}{2.5pt}
  \renewcommand{\arraystretch}{0.94}
  \begin{tabular}{@{}p{0.24\columnwidth}p{0.69\columnwidth}@{}}
    \toprule
    \textbf{Symbol} & \textbf{Definition} \\
    \midrule
    $V$ & Set of all GPU ranks. \\
    $V_a$ & Switch-local group of topology-equivalent GPU ranks, indexed by $a$. \\
    $\mathcal R_{ad}$ & Set of group-level path instances from source group $V_a$ to destination group $V_d$. \\
    $\pi$ & Nonself permutation mapping each source rank $i$ to a distinct destination $\pi(i)$. \\
    $C$ & Channel specifying permutation-defined peer pairs and one concrete communication path per pair. \\
    $E$ & Channel activation obtained by assigning a nonzero lane-volume vector to a channel. \\
    $X(E)$ & Rank-level demand-contribution matrix served by activation $E$. \\
    $M$ & Channel-shape matrix; $M_{ad}$ is the number of lanes from $V_a$ to $V_d$. \\
    $B,\mathcal B[M]$ & Channel family $B$ is a set of group-level path instances whose transfers can execute concurrently without sharing a modeled resource; catalog bucket $\mathcal B[M]$ contains all families that realize shape $M$. \\
    $D^{\rm rem},\widehat D$ & Unserved rank-level demand and its aggregation by ordered rank-group pair. \\
    \bottomrule
  \end{tabular}
  \vspace{-0.5em}
\end{table}

\parab{Channel definition.}
A \emph{channel} $C=(\pi,\mathbf p)$ consists of a permutation $\pi$ over $V$
and a path vector $\mathbf p=(p_i)_{i\in V}$. The mapping $\pi$ assigns each
source rank $i$ to a distinct destination $\pi(i)\ne i$, and
$p_i\in\mathcal P_{i,\pi(i)}$ selects the corresponding concrete path. Because
$\pi$ is a permutation, each channel contains exactly one outgoing and one
incoming lane per rank, preventing same-direction fan-out and fan-in at an
endpoint. A channel specifies only this peer--path pattern.

A channel activation $E=(C,\mathbf x)$ represents an AlltoAllv stage, assigning
a nonnegative integer volume $x_i$ to each lane $i\in V$. All nonzero-volume
lanes execute concurrently in that stage. Specifically,
$X(E)_{i,\pi(i)}=x_i$ for $i\in V$, and all other entries are zero. Thus,
$X(E)$ records the rank-level demand served in that stage.

\parab{Channel feasibility.}
A channel is \emph{feasible} under our model when no two selected paths share a
modeled directed resource:
\begin{equation}
 \sum_{i\in V}\mathbf 1[\ell\in\mathcal L(p_i)]\le1,
 \qquad \forall\ell\in\mathcal U.
 \label{eq:channel-feasibility}
\end{equation}
Opposite directions and parallel physical links are distinct resources. We
conservatively make each shared PCIe link exclusive within a channel because
concurrent use would make path rates traffic-dependent; transfers sharing it
run in separate activations.

\parab{Schedule validity.}
A schedule is an ordered sequence of activations
$\mathcal S=\langle E_1,\ldots,E_K\rangle$, where
$E_k=(C_k,\mathbf x_k)$.  The schedule is \emph{valid} if (i) every $C_k$ is
feasible and (ii) the activations exactly reconstruct the requested demand,
$\sum_{k=1}^{K}X(E_k)=D$.
This equality formalizes the \textbf{\textit{linear combination}} view from
Section~\ref{sec:background}: each activation contributes a channel-shaped
component $X(E_k)$, and these components sum to $D$.

\parab{Optimization objective.}
We estimate schedule completion time under an execution model in which
activations run sequentially and the lanes within each activation run
concurrently. Let $\lambda\ge0$ denote a fixed per-activation overhead. Under
this model, an activation's modeled duration is its fixed overhead plus the
completion time of its slowest lane. For
$E_k=(C_k,\mathbf x_k)$ with permutation $\pi_k$ and paths $\mathbf p_k$,
summing these durations gives
\begin{equation}
 \widehat T(\mathcal S)=
 \sum_{k=1}^{K}\left(
 \lambda+\max_{i\in V}
 \frac{x_{k,i}}{\mathrm{bw}(p_{k,i})}\right),
 \qquad
 \mathcal S^\star\in\arg\min_{\mathcal S\ \mathrm{valid}}
 \widehat T(\mathcal S).
 \label{eq:modeled-completion-objective}
\end{equation}
The objective captures the trade-off between shortening communication tails
and limiting the overhead incurred by additional activations.

\section{\mydesign{} Design}
\label{sec:channel-scheduling}
\begin{figure}[!t]
  \captionsetup{skip=2pt}
  \centering
  \includegraphics[width=\columnwidth]
    {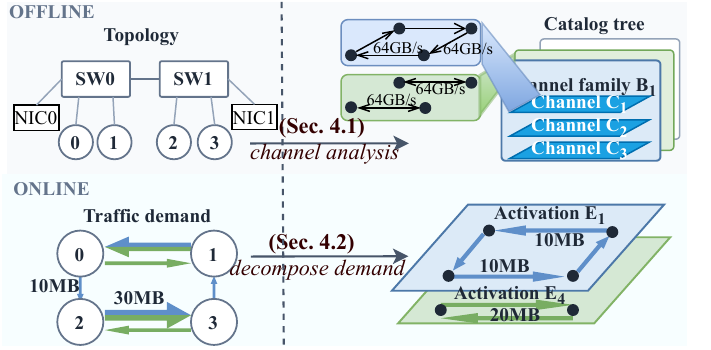}
  \caption{\textbf{\mydesign{}'s workflow.}
  Offline analysis stores compatible channel families in a reusable
  catalog tree; online channel-guided decomposition instantiates concrete
  channels and allocates each invocation's demand across the resulting activations.}
  \label{fig:variopath-overview}
\end{figure}

\parab{Overview.}
\mydesign{} constructs schedules through an offline--online workflow, as shown
in Figure~\ref{fig:variopath-overview}. Offline, it represents sets of
feasible channels as channel families and organizes the channel
families in a reusable catalog tree for the fixed topology. This catalog tree
records feasible choices once, avoiding repeated topology analysis
for each demand. Online, each iteration instantiates candidate channels from
the catalog, assigns lane volumes, and selects the highest-utility activation,
progressively decomposing the demand.

\subsection{Topology-Aware Channel Analysis}
\label{subsec:offline-channels}
\label{subsec:offline-macro}

\begin{figure}[!t]
  \captionsetup{skip=2pt}
  \centering
  \includegraphics[width=\columnwidth]
    {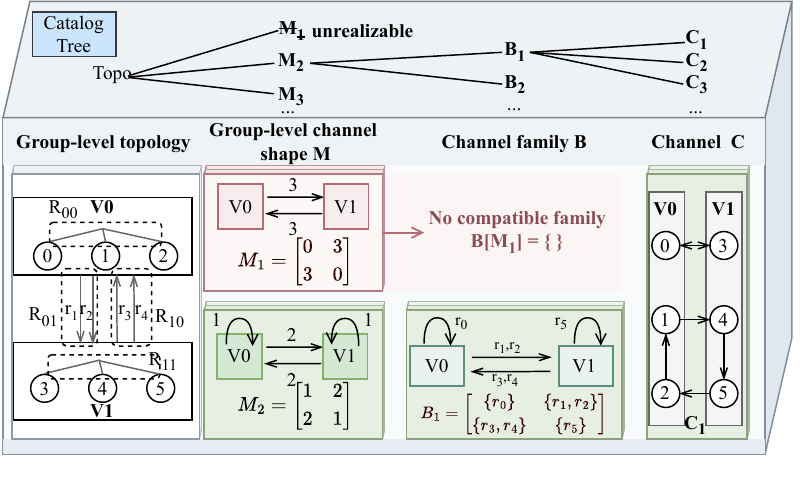}
  \caption{\textbf{Topology-aware construction of the channel-family catalog tree.}
  Balance constraints define admissible shapes, while the group-level topology
  determines which shapes have compatible families. Here $M_1$ is unrealizable
  because $\mathcal B[M_1]=\varnothing$, whereas
  $B_1\in\mathcal B[M_2]$ realizes $M_2$. A one-to-one endpoint binding then
  instantiates $B_1$ as channel $C_1$.}
  \label{fig:offline-catalog-relations}
  \vspace{-1em}
\end{figure}

\parab{Topology abstraction with rank groups and path instances.}
As discussed in Section~\ref{sec:background}, concurrent transfers can contend
on endpoint-local GPU-to-switch links and shared PCIe links, especially
inter-switch links. Because every channel contains exactly one outgoing and one
incoming lane per rank, no endpoint-local link is reused in the same direction.
We therefore model only the remaining conflicts on shared PCIe and NIC-facing
resources in $\mathcal U$. We partition the $N=Gg$ ranks into $G$ switch-local rank groups
$V_a$, each containing $g$ topology-equivalent ranks on the same PCIe switch.
A group-level path instance $r\in\mathcal R_{ad}$ records one possible
shared-resource route from $V_a$ to $V_d$ while omitting endpoint-local
GPU-to-switch attachments. Endpoint binding adds these attachments later; in
Figure~\ref{fig:offline-catalog-relations}, the same instance $r_1$ can thus
connect GPU~0 to GPU~3 or GPU~1 to GPU~4.

\parab{Representing group-pair transfers as channel shapes.}
Because ranks within each group are topology-equivalent, their identities do
not affect the channel's group-level structure. Aggregating a channel's lanes
by ordered rank-group pair therefore preserves its concurrency pattern. We
encode these lane counts in a matrix $M$, called a channel shape. Formally,
$M\in\mathbb Z_{\ge0}^{G\times G}$, where $M_{ad}$ is the number of lanes from
$V_a$ to $V_d$. An admissible shape has balanced rows and columns:
\begin{equation}
 \sum_d M_{ad}=g\quad\forall a,
 \qquad
 \sum_a M_{ad}=g\quad\forall d .
 \label{eq:balanced-m}
 \end{equation}
These constraints allocate exactly $g$ outgoing and $g$ incoming positions to
each group. Assigning the positions one-to-one to its $g$ ranks gives every
rank one outgoing and one incoming lane, as required by the permutation in the
channel definition. A larger $M_{ad}$ assigns more concurrent lanes to
$V_a$-to-$V_d$ traffic, allowing the scheduler to favor group pairs with
greater demand.

\parab{Grouping channels into channel families.}
A shape fixes the number of lanes between each ordered group pair but leaves
their routes unspecified. A channel family $B$ refines a shape $M$ by selecting
group-level path instances while leaving their rank-level endpoints unbound.
Let $B_{ad}(r)\in\{0,1\}$ indicate whether instance
$r\in\mathcal R_{ad}$ is selected. A family realizes $M$ only if it satisfies:
\begin{equation}
 \begin{aligned}
 \sum_{r\in\mathcal R_{ad}}B_{ad}(r)&=M_{ad},
 &&\forall(a,d),\\[-2pt]
 \sum_{a,d}\sum_{r\in\mathcal R_{ad}}
 B_{ad}(r)\mathbf 1[\ell\in\mathcal L(r)]&\le1,
 &&\forall\ell\in\mathcal U .
 \end{aligned}
 \label{eq:family-compatibility}
\end{equation}
The first constraint enforces shape consistency by selecting exactly $M_{ad}$
path instances for every ordered group pair $(a,d)$. The second enforces
resource compatibility by allowing each modeled directed resource
$\ell\in\mathcal U$ to appear in at most one selected instance. Each compatible
family is stored in the bucket $\mathcal B[M]$; hence, $M$ is realizable exactly
when $\mathcal B[M]\ne\varnothing$. Figure~\ref{fig:offline-catalog-relations}
shows that satisfying the balance constraints alone does not guarantee
realizability: $M_1$ has no compatible family, whereas
$B_1\in\mathcal B[M_2]$ realizes $M_2$.

\parab{Instantiating a channel from a family.}
A family fixes compatible path instances but represents multiple rank-level
channels because their endpoints remain unbound. To instantiate one channel,
endpoint binding assigns each selected instance $r\in\mathcal R_{ad}$ a pair
$(i,j)\in\Gamma_{ad}$, where
$\Gamma_{ad}=\{(i,j)\in V_a\times V_d:i\ne j\}$ excludes self-transfers. These
choices are subject to one-to-one source and destination assignments within
each group. Together with the balance constraints, a valid binding uses every
rank exactly once as a source and once as a destination, yielding the
permutation required by Section~\ref{sec:scheduling-model}. Assigning $(i,j)$
turns $r$ into a concrete path
$p\in\mathcal P_{ij}$ by adding its GPU-to-switch attachments. Under the
group-equivalence assumption, these attachments are excluded from
$\mathcal U$ and preserve the modeled footprint and configured rate:
\mbox{$\mathcal L(p)=\mathcal L(r)$} and
\mbox{$\mathrm{bw}(p)=\mathrm{bw}(r)$}.

\begin{algorithm}[t]
\caption{Offline catalog tree construction}
\label{alg:offline-catalog}
\footnotesize
\begin{algorithmic}[1]
\REQUIRE Path-instance sets $\{\mathcal R_{ad}\}$; group size $g$
\ENSURE Pruned channel-family catalog tree $\mathcal B$
\STATE order the instances as $\mathcal R=(r_1,\ldots,r_m)$;
  $\mathcal B\gets\varnothing$
\STATE \textsc{Explore}$(1,\varnothing)$
\STATE $\mathcal B[M]\gets\textsc{PruneFamilies}(\mathcal B[M])$
  for all $M\in\operatorname{dom}(\mathcal B)$
\RETURN $\mathcal B$
\STATE \textbf{procedure} \textsc{Explore}$(t,B)$
  \IF{\textsc{Complete}$(B)$}
    \STATE $M_{ad}\gets|B\cap\mathcal R_{ad}|$ for all $(a,d)$
    \STATE insert $B$ into $\mathcal B[M]$; \textbf{return}
  \ENDIF
  \STATE \textbf{if} $t>m$ or not
    \textsc{SuffixOK}$(t,B)$
    \textbf{ then return}
  \STATE \textsc{Explore}$(t+1,B)$
  \STATE let $r_t\in\mathcal R_{ad}$
  \IF{\textsc{CanAdd}$(r_t,B)$}
    \STATE \textsc{Explore}$(t+1,B\cup\{r_t\})$
  \ENDIF
\end{algorithmic}
\end{algorithm}

\parab{Depth-first construction of the catalog tree.}
Algorithm~\ref{alg:offline-catalog} enumerates candidate path instances in a
fixed order using a pruned binary DFS. At state $(t,B)$, $B$ is the partial
family and $r_t$ the next candidate; the fixed order gives each subset a unique
decision sequence. The exclude branch skips $r_t$, whereas the include branch
is explored only if \textsc{CanAdd} preserves resource disjointness and the
group-degree limits. \textsc{SuffixOK} stops when the remaining instances
cannot fill every group's $g$ outgoing and $g$ incoming slots. Because these
tests reject only incompatible or uncompletable states, all valid families
remain reachable. \textsc{Complete} holds once every group has $g$ selected
outgoing and incoming instances. Then $B$ is valid; its group-pair counts
determine shape $M$, and $B$ is inserted into $\mathcal B[M]$.

\parab{Pruning and search cost.}
Dominance pruning operates at two levels. Before the DFS,
\textsc{PrunePaths} removes a path instance $r$ when an alternative $r'$ for
the same group pair satisfies
$\mathcal L(r')\subseteq\mathcal L(r)$ and
$\mathrm{bw}(r')\ge\mathrm{bw}(r)$, with at least one strict inequality.
Such an alternative uses no additional modeled resource and offers no lower
rate. After the DFS, \mbox{\textsc{PruneFamilies}} removes exact duplicates and
families dominated within the same shape. One family dominates another when
their instances can be paired within every group pair with identical
footprints and no lower rates, with at least one paired instance having a
higher rate. Families with different footprints remain as alternatives because
they expose different resource choices to endpoint binding. These two filters
therefore target different costs: path-level pruning reduces the DFS branching
factor, whereas family-level pruning limits the catalog alternatives exposed
to endpoint binding and online scheduling.

The DFS may visit $O(2^{|\mathcal R|})$ candidate subsets in the worst case,
although resource checks, suffix pruning, and the two dominance filters reduce
the explored and stored choices in practice. The catalog is built once per
machine configuration and reused across invocations, keeping this topology
search off the per-invocation critical path.

\subsection{Channel-Guided Demand Decomposition}
\label{subsec:online-macro}

\begin{figure}[!t]
  \captionsetup{skip=2pt}
  \centering
  \includegraphics[width=\columnwidth]
    {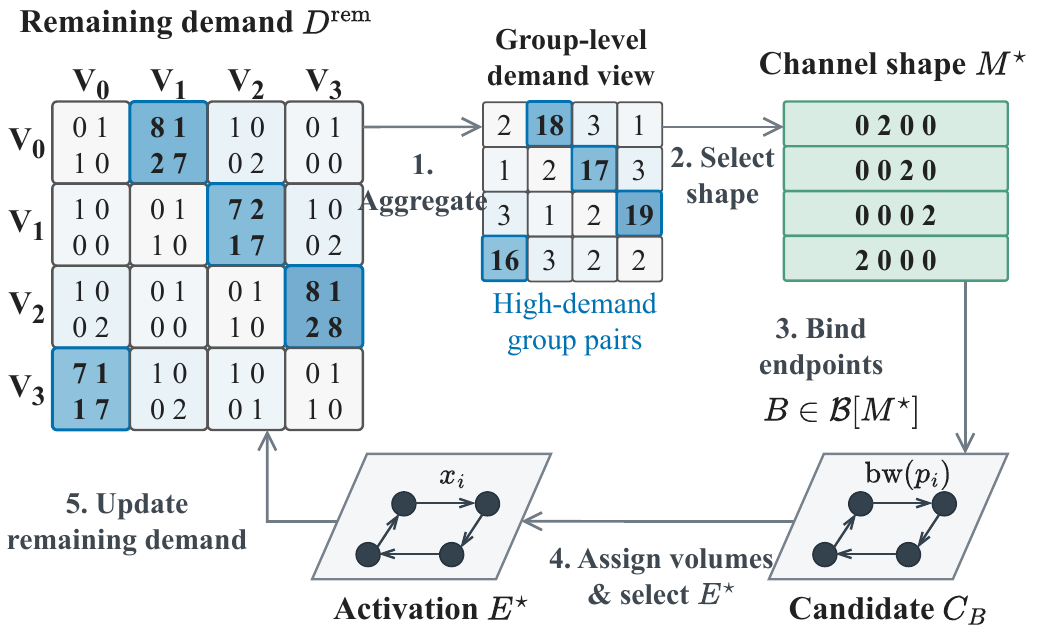}
  \caption{\textbf{One iteration of channel-guided demand decomposition.}
  The group-level view summarizes the remaining demand $D^{\rm rem}$ by
  rank-group pair and guides the selection of channel shape $M^\star$.
  For each family $B\in\mathcal B[M^\star]$, rank-level endpoint binding
  instantiates candidate channel $C_B$. Candidate volumes and utilities
  determine activation $E^\star$, which updates $D^{\rm rem}$.
  Channels are shown schematically.}
  \label{fig:online-search-tree}
  \vspace{-0.55em}
\end{figure}

\parab{Online scheduling procedure.}
Algorithm~\ref{alg:online-scheduler} repeatedly decomposes the remaining demand
into channel activations; Figure~\ref{fig:online-search-tree} illustrates one
iteration. The scheduler initializes $D^{\rm rem}=D$, where $D^{\rm rem}$ is
the demand not yet served. Each iteration traverses the catalog from coarse to
fine. It first aggregates $D^{\rm rem}$ by rank-group pair, scores all channel
shapes, and selects the highest-scoring shape $M^\star$. It then uses the
rank-level demand to bind every family $B\in\mathcal B[M^\star]$ to endpoints,
yielding a candidate channel $C_B$. For each candidate, it assigns lane volumes
and evaluates the full modeled utility $U_B$. The scheduler appends the
highest-utility activation $E^\star$, subtracts its served demand from
$D^{\rm rem}$, and repeats until $D^{\rm rem}=0$. This coarse-to-fine search
narrows the topology choices at the group level while retaining demand-aware
endpoint and volume assignment.

\begin{algorithm}[t]
\caption{Online channel-guided demand decomposition}
\label{alg:online-scheduler}
\footnotesize
\begin{algorithmic}[1]
\REQUIRE Feasible demand $D$; complete catalog tree $\mathcal B$;
  $\tau>0$, $\lambda\ge0$;
  Alternating Algorithm (AA) hyperparameters
  $N_{\rm start},I_{\rm AA}>0$
\ENSURE Schedule $\mathcal S$ with $\sum_{E\in\mathcal S}X(E)=D$
  \STATE $D^{\rm rem}\gets D$; $\mathcal S\gets\langle\rangle$
\WHILE{$\lVert D^{\rm rem}\rVert_1>0$}
  \STATE $\widehat D_{ad}\gets
    \sum_{i\in V_a}\sum_{j\in V_d}D^{\rm rem}_{ij}$ for all $(a,d)$
  \STATE $M^\star\gets
    \arg\max_{M\in\operatorname{dom}(\mathcal B)}
      \sum_{a,d}\widehat D_{ad}M_{ad}$
  \FOR{$B\in\mathcal B[M^\star]$}
    \STATE $C_B\gets\textsc{MultiStartAA}
      (B,D^{\rm rem},\tau,N_{\rm start},I_{\rm AA})$
    \STATE $(\mathbf x_B,U_B)\gets
      \textsc{Evaluate}(C_B,D^{\rm rem},\tau,\lambda)$
  \ENDFOR
  \STATE $B^\star\gets
    \arg\max_{B\in\mathcal B[M^\star]}U_B$
  \STATE $E^\star\gets(C_{B^\star},\mathbf x_{B^\star})$
  \STATE $\mathcal S\gets\mathcal S\mathbin{\|}\langle E^\star\rangle$
  \STATE $D^{\rm rem}\gets D^{\rm rem}-X(E^\star)$
\ENDWHILE
\RETURN $\mathcal S$
\end{algorithmic}
\end{algorithm}

\parab{Step 1: Bottleneck-aware channel-shape selection.}
A channel shape specifies group-pair lane counts, so the scheduler aggregates
the remaining demand at the same granularity:
\begin{equation}
 \widehat D_{ad}
 =\sum_{i\in V_a}\sum_{j\in V_d}D^{\rm rem}_{ij}.
 \label{eq:group-demand}
\end{equation}
A group pair with larger $\widehat D_{ad}$ has more residual bytes to drain and
is more likely to determine the completion tail if assigned insufficient
concurrency. We therefore use $\widehat D_{ad}$ as its current bottleneck
pressure. Because Equation~\ref{eq:balanced-m} gives every admissible shape the
same total lane budget, allocating more lanes to one group pair necessarily
reduces the concurrency available to others. We score each shape by
\begin{equation}
 s(M)=\sum_{a,d}\widehat D_{ad}M_{ad},
 \qquad
 M^\star=\arg\max_{M\in\operatorname{dom}(\mathcal B)}s(M).
 \label{eq:best-m}
\end{equation}
\begin{figure}[!t]
  \captionsetup{skip=2pt}
  \centering
  \includegraphics[width=\columnwidth]
    {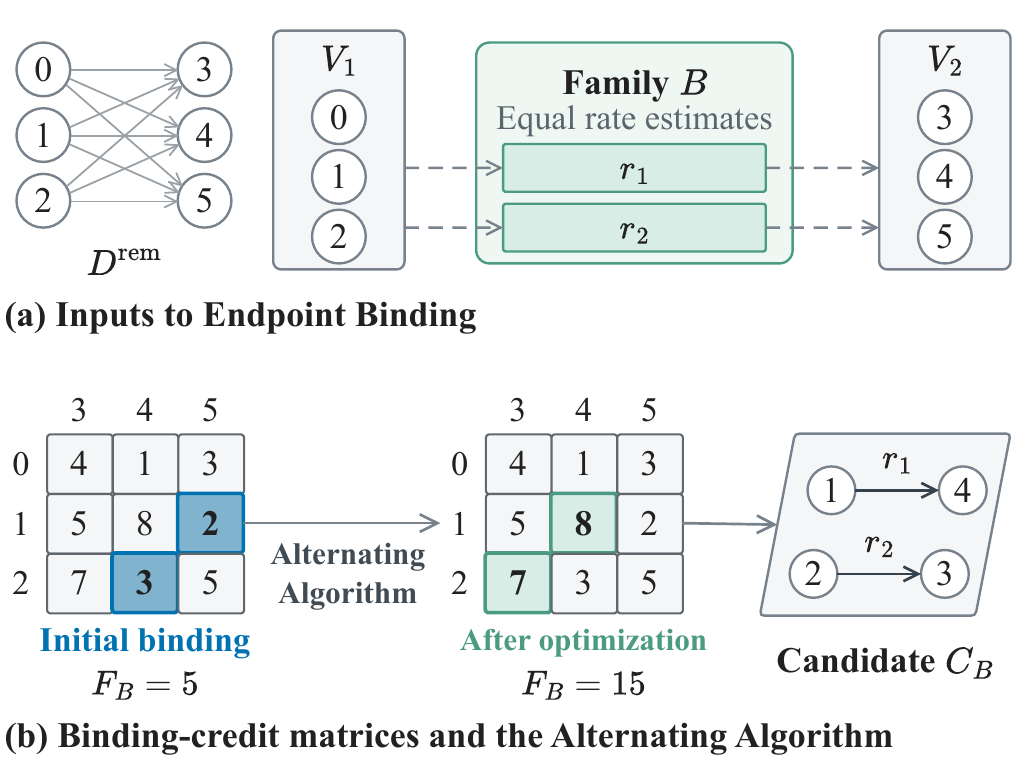}
  \caption{\textbf{Endpoint binding with the Alternating Algorithm.}
  (a) Family $B$ provides two unbound, equal-rate path instances from $V_1$
  to $V_2$. (b) Colored entries in the credit matrices $h_{ijr}$ mark the
  current endpoint assignments. Alternating source and destination optimization
  raises $F_B$ from 5 to 15 and produces candidate channel $C_B$.
  Other family slots are omitted for clarity.}
  \label{fig:candidate-instantiation}
  \vspace{-0.45em}
\end{figure}

The objective directs the fixed concurrency budget toward high-pressure pairs,
while restricting the search to $\operatorname{dom}(\mathcal B)$ guarantees
topology feasibility. Recomputing the pressures and $M^\star$ after each update
to $D^{\rm rem}$ lets the scheduler follow the evolving communication
bottleneck.

\parab{Step 2: Binding paths to form candidate channels.}
For each family $B$ under $M^\star$, the scheduler binds its path instances
one-to-one to rank-level GPU pairs, producing a candidate channel $C_B$.
Let $\tau$ be a fixed invocation-wide time quantum, and let $s_B(r)$ and
$d_B(r)$ denote $r$'s assigned endpoints.
For $r\in\mathcal R_{ad}$ and a valid pair $(i,j)\in\Gamma_{ad}$, define the
useful-rate credit
\begin{equation}
 h_{ijr}
 =\min\!\left\{\frac{D^{\rm rem}_{ij}}{\tau},\mathrm{bw}(r)\right\},
 \label{eq:binding-credit}
\end{equation}
with zero credit outside $\Gamma_{ad}$. The two terms are the demand rate
needed within one quantum and $r$'s supported rate, so their minimum measures
the useful assignment rate. We score a complete binding by
\begin{equation}
 F_B(s_B,d_B)
 =\sum_{r\in B}h_{s_B(r),d_B(r),r}.
 \label{eq:binding-score}
\end{equation}
Thus, $\tau F_B$ is its provisional useful byte volume.

Because each credit jointly depends on both endpoints,
optimizing both one-to-one assignments yields a bilinear assignment problem
(BAP)~\cite{custic2017bilinear}. We search for a high-scoring binding with the
Alternating Algorithm (AA)~\cite{sokol2020bilinear}.
Fixing either endpoint assignment reduces the remaining optimization to a
standard assignment problem, so AA alternately optimizes the source and
destination assignments. Each update solves one
$g\times g$ assignment problem per rank group without decreasing $F_B$. A run
ends when neither update improves the score or the sweep limit is reached.
Figure~\ref{fig:candidate-instantiation} illustrates this process, where the
displayed binding credit rises from 5 to 15.

To reduce initialization sensitivity, \textsc{MultiStartAA} runs AA from
$N_{\rm start}$ feasible initial endpoint bindings. Each run stops after
$I_{\rm AA}$ sweeps or when a complete sweep does not improve $F_B$. The
procedure returns the highest-scoring channel $C_B=(\pi_B,\mathbf p_B)$ for
evaluation in Step~3.

\parab{Step 3: Allocating volumes and selecting an activation.}
For candidate $C_B$, \textsc{Evaluate} assigns each lane the largest volume
allowed by both its path capacity within $\tau$ and its remaining demand:
\begin{equation}
 x_{B,i}
 =\min\!\left\{\tau\mathrm{bw}(p_{B,i}),
                 D^{\rm rem}_{i,\pi_B(i)}\right\}.
 \label{eq:candidate-volume}
\end{equation}
We compare the resulting activations by utility $U_B$: useful bytes divided by
modeled duration, including the slowest-lane time and per-activation overhead
from Section~\ref{sec:scheduling-model}. We define $U_B=0$ for a candidate with
$\sum_i x_{B,i}=0$; otherwise,
\begin{equation}
 U_B
 =
 \frac{
   \sum_{i\in V}x_{B,i}
 }{
   \lambda
   +
   \max_{i\in V}
   \dfrac{x_{B,i}}{\mathrm{bw}(p_{B,i})}
 }.
 \label{eq:candidate-utility}
\end{equation}
The candidate with the largest $U_B$ supplies the next activation $E^\star$.
For feasible nonzero demand and the complete catalog produced by
Algorithm~\ref{alg:offline-catalog}, a positive-scoring $M^\star$ admits at
least one positive-volume candidate. Thus, every iteration reduces
$D^{\rm rem}$.

\parab{Planning complexity.}
Let $J=|\operatorname{dom}(\mathcal B)|$ and
$H^\star=|\mathcal B[M^\star]|$. Aggregating demand and scoring all shapes cost
$O(N^2+JG^2)$ per iteration. Each AA sweep solves $2G$ assignments of size
$g\times g$; with $N_{\rm start}$ starts and at most $I_{\rm AA}$ sweeps,
binding and evaluating all families cost
$O\!\left(H^\star(N_{\rm start}I_{\rm AA}Gg^3+N)\right)$. Thus, $K$
activations cost
$O\!\left(K[N^2+JG^2+H^\star(N_{\rm start}I_{\rm AA}Gg^3+N)]\right)$;
Section~\ref{subsec:planning-overhead} measures the resulting CPU overhead.

\begin{table*}[t]
  \centering
  \footnotesize
  \setlength{\tabcolsep}{3pt}
  \renewcommand{\arraystretch}{1.12}
  \begin{tabular*}{\textwidth}{@{\extracolsep{\fill}}lcccccccc@{}}
    \toprule
    Name & GPU & \makecell{\#GPUs\\per node} & \makecell{PCIe\\generation}
      & \makecell{PCIe BW\\(GB/s)} & NIC & \makecell{\# NIC\\per node}
      & \makecell{NIC BW/node\\(Gb/s)} & \makecell{PCIe\\topology} \\
    \midrule
    R5080-CX8 & RTX 5080 & 8 & Gen5 & 64
      & ConnectX-8 & 4 & 3200 & Tree \\
    R5080-BF3 & RTX 5080 & 8 & Gen5 & 64
      & BF3-mini & 4 & 1600 & Tree \\
    R5090-BF3 & RTX 5090 & 16 & Gen5 & 64
      & BF3-mini & 8 & 3200 & Ring \\
    L20-CX7 & L20 & 16 & Gen4 & 32
      & ConnectX-7 & 8 & 3200 & Tree \\
    \bottomrule
  \end{tabular*}
  \caption{Evaluation platforms. PCIe generation and bandwidth describe each
  GPU's PCIe version and nominal one-way capacity. GPU/NIC counts and aggregate NIC
  bandwidth are per node.}
  \label{tab:eval_testbeds}
\end{table*}

\section{Implementation}
\label{sec:implementation}
\mydesign{} comprises about 3,000 lines of C++ for the offline and online
schedulers and 5,000 lines of C++/CUDA for its runtime communication kernels.
\mydesign{} implements all inter-node communication with NCCL GPU-Initiated
Networking (GIN)~\cite{NCCL} over the GDAKI backend. Overall, the implementation
combines topology-aware path profiling with device-resident data movement and
synchronization, specializes transfers for sparse expert traffic and small
messages, and adapts execution granularity to each demand.

\parab{Topology, path, and rate model.}
At setup, we reconstruct the topology from the GPU/HCA PCIe hierarchy and
socket affinity, enumerate supported direct-P2P and NIC paths, and
record their directed shared-resource footprints and bandwidths. Using this
information, we construct the group-level path instances described in
Section~\ref{subsec:offline-channels}. The runtime loads the resulting profile
once and reuses it across calls.

\parab{Device-resident buffer layout.}
AlltoAllv may scatter each destination's data across noncontiguous user-buffer
regions. Sending fragments separately creates small transfers and underutilizes
links. A customized packing kernel groups elements by destination rank and
packs each peer's metadata and payload contiguously into the symmetric send
buffer; the receive kernel unpacks them directly into the output layout.
This fused layout conversion eliminates a separate memory-reordering pass and
avoids fine-grained transfers.

To avoid the high cost of a global barrier, measured at approximately
20~$\mu$s on our 32-GPU RTX~5090 testbed, we use double buffering to eliminate
the global barriers otherwise required before and after resetting signals. We
implement this scheme with two symmetric-memory slots, each containing
independent payloads, signal counts, indices, and other invocation-related state
variables. In practice, CUDA Graphs capture and replay
these communication operations. To avoid host--device state
inconsistencies across replays, the counters, slot indices, and related state
reside in device memory rather than host memory.

\parab{Efficient sparse data movement for expert parallelism.}
In sparse AlltoAllv workloads, source--destination rank pairs with zero demand
require neither data transfer nor completion waits. For a destination rank
hosting $L$ local experts, each active source--destination pair uses a 32-bit
control word: the lower $L$ bits indicate which local experts receive data,
while the remaining usable high-order bits encode the exact token count for a
designated local expert. Each local expert is mapped to a queue pair (QP),
using a dedicated QP when resources permit and sharing QPs round-robin
otherwise. The sender posts the designated expert's payload followed by the
control word on the same QP, so NIC-level in-order execution guarantees
data-before-control ordering without an additional fence. The receiver uses the
bitmap to skip inactive experts and obtains active-expert counts from GIN
signals, avoiding separate count transfers and output-count atomics for empty
groups.

\parab{Low-latency protocol.}
GIN signal mode appends an RDMA atomic notification after the payload, incurring
additional notification overhead. NCCL LL avoids the separate notification by
embedding 8 bytes of flags alongside every 8 bytes of payload in each 16-byte
FIFO line, reducing payload utilization to 50\%. To achieve ultra-low latency
without this overhead, we design a specialized protocol. Receivers initialize
idle buffers to a sentinel value (e.g., negative zero), and senders issue
unsignaled puts. Receivers poll 16-byte vectors until no sentinel remains. In a
1~KiB protocol microbenchmark on two RTX 5090 servers, P50 latency is
7.872~$\mu$s;
GIN signal and prepacked NCCL LL take 12.928 and 9.726~$\mu$s, respectively.
To avoid split PCIe TLPs, we pad the receive-slot stride so that each token's
receive address is 128-byte aligned.

\parab{Global execution granularity.}
Each invocation begins with a target round count $K_{\mathrm{target}}$, the
approximate number of scheduling rounds over which the current demand should be
drained. Given $K_{\mathrm{target}}$, the current demand, and estimated path
rates, the runtime derives a global time quantum $\tau$ so that the transfer is
expected to complete in roughly $K_{\mathrm{target}}$ rounds. A smaller
$K_{\mathrm{target}}$ therefore yields a larger $\tau$ and fewer rounds,
reducing scheduling and synchronization overhead, whereas a larger
$K_{\mathrm{target}}$ yields a smaller $\tau$ and enables finer-grained
adaptation of peers and paths as demand drains. In practice, for an $N$-rank
invocation, we set $K_{\mathrm{target}}$ between $N$ and $3N$.

\begin{figure*}[t!]
  \captionsetup{skip=1pt}
  \centering
  \includegraphics[width=.95\textwidth]{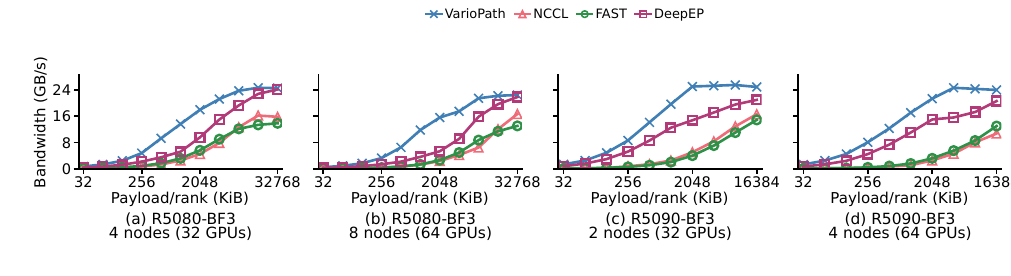}
  \caption{Random-demand AlltoAllv bandwidth on 32- and 64-GPU deployments.
  Bandwidth is per-rank payload divided by p50 completion time.}
  \label{fig:alltoallv-fast}
  \vspace{-0.35em}
\end{figure*}

\begin{figure*}[t!]
  \captionsetup{skip=1pt}
  \centering
  \includegraphics[width=.265\textwidth]{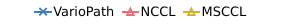}
  \par\vspace{-.35em}
  \subfigure[R5080-CX8, 1 node (8 GPUs)]{%
    \includegraphics[width=.226\textwidth]{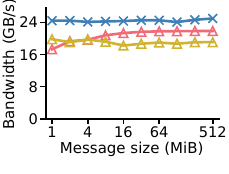}}%
  \hfill
  \subfigure[R5080-BF3, 1 node (8 GPUs)]{%
    \includegraphics[width=.226\textwidth]{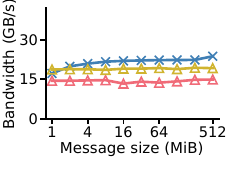}}%
  \hfill
  \subfigure[R5090-BF3, 1 node (16 GPUs)]{%
    \includegraphics[width=.226\textwidth]{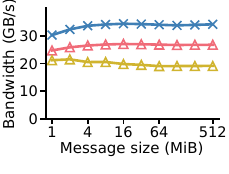}}%
  \hfill
  \subfigure[L20-CX7, 1 node (16 GPUs)]{%
    \includegraphics[width=.226\textwidth]{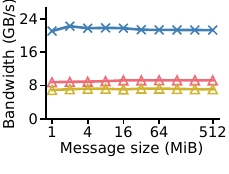}}%
  \par\vspace{-.35em}
  \subfigure[R5080-CX8, 2 nodes (16 GPUs)]{%
    \includegraphics[width=.226\textwidth]{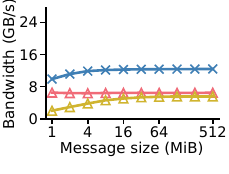}}%
  \hfill
  \subfigure[R5080-BF3, 2 nodes (16 GPUs)]{%
    \includegraphics[width=.226\textwidth]{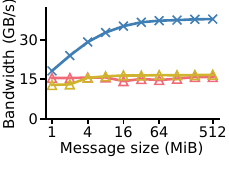}}%
  \hfill
  \subfigure[R5090-BF3, 2 nodes (32 GPUs)]{%
    \includegraphics[width=.226\textwidth]{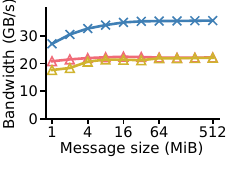}}%
  \hfill
  \subfigure[L20-CX7, 2 nodes (32 GPUs)]{%
    \includegraphics[width=.226\textwidth]{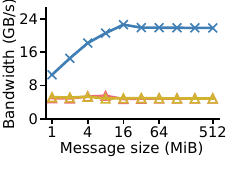}}
  \caption{AlltoAll bandwidth on one- and two-node deployments.
  Bandwidth is per-rank payload divided by completion time.}
  \label{fig:large-alltoall}
  \vspace{-0.35em}
\end{figure*}

\flushbottom
\section{Evaluation} \label{sec:6}
\setlength{\textfloatsep}{8pt plus 1pt minus 1pt}
\setlength{\floatsep}{6pt plus 1pt minus 1pt}

We evaluate \mydesign{} to answer four key questions:
\begin{itemize}
  \setlength{\itemsep}{0pt}
  \setlength{\parsep}{0pt}
  \setlength{\topsep}{2pt}
  \setlength{\partopsep}{0pt}
  \item How does \mydesign{} compare with state-of-the-art systems across
  collective workloads, transfer sizes, and deployment scales
  (\S\ref{subsec:eval-alltoallv})?
  \item What end-to-end latency reductions does \mydesign{} provide for MoE
  and sequence-parallel inference workloads (\S\ref{subsec:e2e-model})?
  \item How closely does \mydesign{}'s channel-level peer--path co-design
  approach the full MILP optimum
  (\S\ref{subsec:scheduling-sensitivity})?
  \item How scalable are \mydesign{}'s offline catalog and online planning
  (\S\ref{subsec:planning-overhead})?
\end{itemize}

\parab{Testbeds.}
Table~\ref{tab:eval_testbeds} lists the GPUs, PCIe generations and bandwidths,
NICs, and topology of our four platforms. Our testbed comprises 32 R5080-CX8
nodes, 8 R5080-BF3 nodes, 4 R5090-BF3 nodes, and 2 L20-CX7 nodes.
The nodes are interconnected through a two-tier leaf--spine network.

\parab{Workloads.}
We evaluate \mydesign{} with random-demand AlltoAllv, skew AlltoAllv, AlltoAll,
and end-to-end inference. In AlltoAll, every rank sends the same volume to every
other rank. Random-demand AlltoAllv uses demand matrices derived from randomized
top-$k$ MoE routing over 256 experts, with $k=8$. Following
FAST~\cite{fastalltoall}, we generate skew AlltoAllv by drawing pairwise volumes
from a Zipfian distribution.
Our end-to-end workloads cover expert-parallel Qwen3 prefill and
sequence-parallel Wan2.1 generation.
A separate 256-GPU experiment evaluates \mydesign{}'s large-scale scalability.

\parab{Metrics.}
Our primary communication metric is \emph{algorithmic bandwidth}, computed as
the total transferred payload divided by the product of the number of ranks
and completion time~\cite{fastalltoall}. Higher is better. Because skew
AlltoAllv can make ranks finish at different times, invocation
completion time is the maximum communication time across ranks.
Aggregate comparisons use geometric-mean speedup over matched points.
End-to-end experiments report latency. Planning experiments report online CPU
time per emitted activation and the ratio of total online-planning time to
communication time; lower is better for both.

\parab{Baselines.}
Across the communication and end-to-end experiments, we compare \mydesign{}
with NCCL~\cite{NCCL}, MSCCL~\cite{cowan2023mscclang},
FAST~\cite{fastalltoall}, and DeepEP~\cite{deepep}\footnote{We evaluate
DeepEP V2 from its
\href{https://github.com/deepseek-ai/DeepEP}{official GitHub repository}.}
MSCCL executes user-defined schedules that control transfer ordering and
concurrency. FAST
is built on NVSHMEM and rebalances skewed traffic within each node before
balanced one-to-one inter-node transfers. DeepEP provides specialized MoE
dispatch and combine kernels over NCCL GIN.

\subsection{Collective Communication Performance}
\label{subsec:eval-alltoallv}

We first evaluate random-demand AlltoAllv across transfer sizes and scales,
then AlltoAll and skew AlltoAllv, and finally large-scale random-demand
AlltoAllv on 256 GPUs.
These experiments test whether \mydesign{}'s benefits persist across payload
sizes, deployment scales, and traffic distributions.

\parab{Random-demand AlltoAllv.}
We vary the per-rank payload on 32- and 64-GPU deployments with different local
PCIe organizations. R5080-BF3 uses four and eight 8-GPU nodes, whereas
R5090-BF3 uses two and four 16-GPU nodes.
Figure~\ref{fig:alltoallv-fast} shows that
\mydesign{} achieves the highest bandwidth on all 42 matched test points. Its
geometric-mean speedup over DeepEP, the closest baseline, ranges from
$1.55\times$ to $2.00\times$. Over FAST, it ranges from $3.49\times$ to
$9.41\times$, and over NCCL from $3.73\times$ to $10.61\times$. By jointly
selecting each channel's peer matching and physical paths, \mydesign{} avoids
concentrating concurrent flows on shared PCIe links and maps chunks onto
concurrently usable PCIe and NIC paths.

To examine scaling, we fix the per-rank payload while increasing the number of
GPUs. At 32~MiB on R5080-BF3, scaling from 32 to 64 GPUs
reduces \mydesign{} bandwidth by 8.6\% and FAST bandwidth by 5.7\%, while
\mydesign{} remains $1.72\times$ faster. At 16~MiB per rank on R5090-BF3, the
corresponding reductions are 3.6\% and 12.6\%. Thus, \mydesign{} retains its
bandwidth advantage at 64 GPUs on both platforms by re-selecting channels
across rounds as individual transfers complete, instead of leaving the
additional flows pinned to a small set of shared links.

\parab{AlltoAll.}
\label{subsec:uniform-alltoall}
\mydesign{}'s advantage persists under uniform traffic distributions. Across the 80
configurations in Figure~\ref{fig:large-alltoall}, it achieves geometric-mean
speedups of $1.82\times$ over NCCL and $1.99\times$ over MSCCL. It
outperforms NCCL in every configuration and MSCCL in all but the
1-MiB, single-node R5080-BF3 case. NCCL fixes transport paths at communicator
setup, and MSCCL inherits those paths. Neither baseline can adapt an active
pair's physical path to the current stage. \mydesign{} instead adapts paths by
stage, distributing uniform traffic across usable PCIe and NIC paths and
avoiding concentration on shared links. The result confirms that this benefit
is not limited to skewed traffic.

\begin{figure*}[t!]
  \captionsetup{skip=1pt}
  \centering
  \includegraphics[width=.95\textwidth]{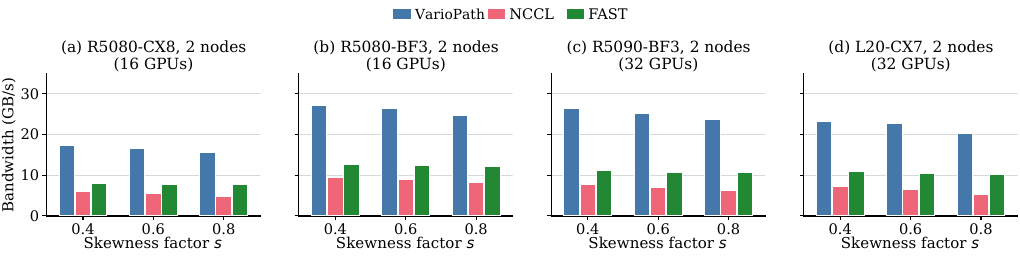}
  \caption{Skew AlltoAllv bandwidth on four two-node
  deployments. Each bar averages five Zipfian workload seeds.}
  \label{fig:alltoallv-skew-platforms}
  \vspace{-0.35em}
\end{figure*}

\begin{figure}[!t]
  \captionsetup{skip=1pt}
  \centering
  \includegraphics[width=.95\columnwidth]
    {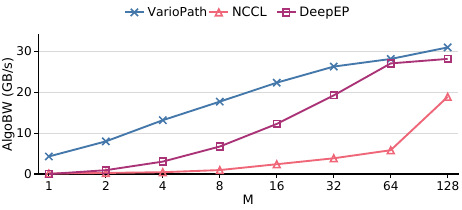}
  \caption{Large-scale random-demand AlltoAllv on 256 R5080-CX8 GPUs. $M$ is
  the number of
  input tokens per rank; hidden dimension is 7168 and top-$k$ is 8.}
  \label{fig:large-scale-256g}
  \vspace{-0.35em}
\end{figure}

\begin{figure}[!t]
  \captionsetup{skip=1pt}
  \centering
  \includegraphics[width=.95\columnwidth]{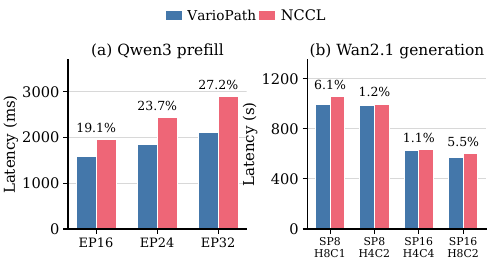}
  \caption{End-to-end latency for (a) Qwen3 prefill and (b) Wan2.1
  generation. Labels report latency reduction from NCCL to \mydesign{}.}
  \label{fig:e2e-model-comparison}
  \vspace{-0.35em}
\end{figure}

\parab{Skew AlltoAllv.}
We increase the Zipf skewness factor from 0.4 to 0.8 on all four two-node
deployments. Larger factors concentrate more traffic in fewer rank pairs,
making the demand harder to distribute across concurrent paths.
Figure~\ref{fig:alltoallv-skew-platforms} shows that \mydesign{} remains
fastest in all 12 deployment--skew configurations, with a geometric-mean
speedup of $2.16\times$ over FAST, ranging from $1.99\times$ to $2.39\times$
across platforms. From $s=0.4$ to 0.8, its bandwidth
drops 9.4--12.3\%, versus NCCL's 15.2--27.0\%. FAST declines less
(2.0--6.0\%), but starts from a substantially lower bandwidth and remains
slower throughout the sweep. FAST's node-local rebalancing is less effective
on PCIe systems because its additional intra-node forwarding competes with
GPU-to-NIC transfers for shared PCIe bandwidth. Balancing NIC traffic can
therefore leave the PCIe bottleneck unresolved or intensify it.
By re-scoring families and rebinding peers, paths, and chunks after each
activation, \mydesign{} can redirect outstanding traffic to newly available
path capacity.

\parab{Large-scale experiment.}
We further evaluate \mydesign{} on 32 R5080-CX8 nodes with 256 GPUs using a
random-demand AlltoAllv workload generated by top-$k$ MoE routing with $k=8$
and hidden dimension 7168.
Figure~\ref{fig:large-scale-256g} sweeps the per-rank input-token count $M$
from 1 to 128. \mydesign{} achieves the highest algorithmic bandwidth at all
eight values of $M$ and reaches 30.91~GB/s at $M=128$, compared with
28.11~GB/s for DeepEP and 18.85~GB/s for NCCL. This result shows that
demand-aware channel scheduling continues to expose usable PCIe and NIC
capacity when the deployment grows to hundreds of GPUs.

\subsection{End-to-End Inference Performance}
\label{subsec:e2e-model}

\begin{figure}[!t]
  \captionsetup{skip=1pt}
  \centering
  \includegraphics[width=.95\columnwidth]
    {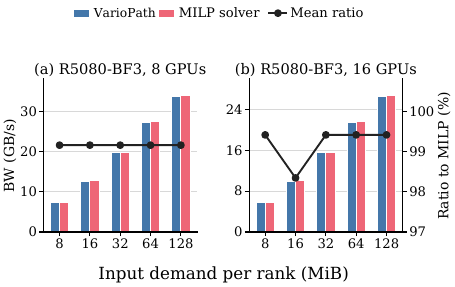}
  \caption{Scheduler quality against a full MILP optimum. Bars show mean
  per-rank bandwidth, and the line shows the mean \mydesign{}/MILP ratio over
  five demand matrices per size.}
  \label{fig:scheduler-quality}
  \vspace{-0.35em}
\end{figure}

\begin{table}[t]
  \centering
  \caption{Peer--path co-design ablation on two-node R5080-BF3.
  Values are completion time normalized to demand-aware peers with dynamic paths.}
  \label{tab:scheduler-ablation}
  \scriptsize
  \setlength{\tabcolsep}{5.0pt}
  \begin{tabular}{@{}lcc@{}}
    \toprule
    & Fixed nearest path & Dynamic path \\
    \midrule
    Fixed peer rotation & 2.84$\times$ & 2.00$\times$ \\
    Demand-aware peers & 2.03$\times$ & \textbf{1.00$\times$} \\
    \bottomrule
  \end{tabular}
  \vspace{-0.35em}
\end{table}

\begin{table}[t]
  \centering
  \caption{Catalog tree scale and online planning overhead. Catalog tree size
  is the estimated compact serialized payload. Online is planning time per
  emitted activation; O/C is total online-planning time divided by
  communication time.}
  \label{tab:catalog-overhead}
  \scriptsize
  \setlength{\tabcolsep}{1.25pt}
  \renewcommand{\arraystretch}{0.96}
  \resizebox{\columnwidth}{!}{%
    \begin{tabular}{@{}lrrrrrr@{}}
      \toprule
      Deployment &
      \makecell[r]{\#\\Shapes} &
      \makecell[r]{\#\\Families} &
      \makecell[r]{Catalog\\tree\\(KiB)} &
      \makecell[r]{Offline\\(s)} &
      \makecell[r]{Online\\($\mu$s/act.)} &
      \makecell[r]{O/C\\(\%)} \\
      \midrule
R5080-BF3--8 GPUs & 3 & 6 & 1.7 & 0.800 & 3.2 & 0.0715 \\
R5080-BF3--16 GPUs & 41 & 438 & 132.0 & 4.950 & 20.6 & 0.7554 \\
R5080-CX8--256 GPUs & 1616 & 2140 & 172682.6 & 5673.603 & 78.12 & 4.1241 \\
R5090-BF3--16 GPUs & 115 & 248 & 123.0 & 15.520 & 8.3 & 0.3249 \\
R5090-BF3--32 GPUs & 308 & 389 & 806.7 & 99.250 & 21.1 & 1.1051 \\
L20-CX7--16 GPUs & 92 & 270 & 121.3 & 11.720 & 6.8 & 0.1476 \\
L20-CX7--32 GPUs & 352 & 462 & 905.0 & 86.940 & 20.8 & 0.7504 \\
\bottomrule

    \end{tabular}%
  }
  \vspace{-0.35em}
\end{table}

We replace NCCL-based AlltoAllv communication with \mydesign{} in uncached
Qwen3--30B--A3B prefill~\cite{sglang,qwen3tech} and Wan2.1 video
generation~\cite{wan2025}.
Qwen runs on two to four R5080-BF3 nodes with EP16, EP24, and EP32; each trial
prefills 8,192 input tokens and generates one token, and each point is the
client median of 15 requests. Wan2.1 runs on 8 and 16 L20 GPUs and generates a
five-second BF16 $1280\!\times\!720$ video with sequence length 75,600, 40
attention heads, and head dimension 128. In Figure~\ref{fig:e2e-model-comparison},
H and C denote the head- and context-parallel factors within the stated
sequence-parallel degree.

For Qwen3--30B--A3B, replacing the exchange with \mydesign{} reduces prefill
latency by 19.1--27.2\% across EP16, EP24, and EP32. The reduction increases
with the expert-parallel degree because expert dispatch and combine lie on the
prefill critical path, and larger degrees expose more cross-GPU and cross-node
traffic to demand-aware channel scheduling.

For Wan2.1, the speedup ranges from $1.011\times$ to $1.066\times$ across the
four sequence-parallel layouts, with the largest gains at H8C1 and H8C2.
At a fixed sequence-parallel degree, comparing H8C1 with H4C2 and H8C2 with
H4C4 shows how the head/context decomposition reshapes rank-pair demand.
Thus, the end-to-end benefit depends on workload and layout, not the
sequence-parallel degree alone.

\subsection{Design Validation}
\label{subsec:scheduling-sensitivity}

We first use an ablation study to quantify the contribution of joint
peer--path selection, and then compare \mydesign{} with a full MILP optimum to
measure the optimality gap of its online scheduler.

\parab{Peer/path ablation.}
This ablation tests the offline channel hypothesis that high-throughput
execution requires jointly retaining peer matchings and path choices that avoid
modeled shared-resource contention. To isolate this effect, we keep all other
system components unchanged and vary only the eligible channel set. We replay
the same five skew AlltoAllv demand matrices with $s=0.6$ on two-node
R5080-BF3. Fixed peers
restrict each stage to a cyclic peer rotation, while fixed paths use each
pair's preassigned nearest path; the other dimension remains adaptive.
Table~\ref{tab:scheduler-ablation} shows that fixing either dimension
approximately doubles completion time, while fixing both increases it to
$2.84\times$ that of the unrestricted design.
Thus, peer--path co-design is necessary to retain the unrestricted design's
performance on this workload.
With only dynamic paths, a cyclic peer rotation can still select flows that
compete for the same shared links; with only demand-aware peers, preassigned
paths cannot redirect those flows to idle capacity. The similar penalties of
the two one-sided variants show that neither decision substitutes for the
other.

\parab{Scheduler quality.}
To evaluate scheduling quality, we compare \mydesign{} with a full MILP
optimum on one- and two-node R5080-BF3 deployments. We use per-rank input
demands of 8, 16, 32, 64, and 128~MiB and evaluate five independently generated
dense Zipf demand matrices for each topology and input size, giving 50 complete
skew AlltoAllv calls. Following the modeling approach of
TE-CCL~\cite{liu2024teccl}, the MILP jointly optimizes concrete peer matchings,
physical paths, transfer sizes, and their execution order over the entire
call. It searches the complete modeled path space, and every instance reaches
a solver-certified optimum. Figure~\ref{fig:scheduler-quality} reports
the algorithmic bandwidth of both methods.
Across all 50 instances, \mydesign{} attains 99.18\% of the MILP-optimal
bandwidth on average, corresponding to an average optimality gap of 0.82\%.
After averaging within each topology--input-size configuration, the ratio
ranges from 98.34\% to 99.41\%, with a minimum of 91.92\% among individual
instances.

\subsection{Catalog Tree Scale and Planning Overhead}
\label{subsec:planning-overhead}

\parab{Catalog tree size.}
Table~\ref{tab:catalog-overhead} reports offline catalog construction across
deployments from 8 to 256 GPUs. The retained group-level trees contain
3--1,616 shapes and 6--2,140 families, occupy between 1.7~KiB and
168.6~MiB, and take between 0.8~s and 94.6~min to construct. Construction
applies
compatibility filtering, structural deduplication, and rate-dominance pruning
without expanding concrete rank bindings, so one family represents many
concrete channels. This one-time cost is amortized across demands on the same
topology. Equal-size R5090-BF3 and L20-CX7 deployments retain different numbers
of shapes and families, showing that catalog scale depends on topology as well
as GPU count.

\parab{Planning cost.}
We average online decomposition over 100 1-GiB-per-source demands. Planning
takes 3.2--78.12~$\mu$s per activation; total
planning time accounts for 0.0715--4.1241\% of communication time (O/C in
Table~\ref{tab:catalog-overhead}). Even at 256 GPUs, planning takes
78.12~$\mu$s per activation, while total planning time accounts for 4.1241\%
of communication time. Encoding topology feasibility offline
limits online work to family search, endpoint binding, and transfer-size
assignment, enabling per-invocation adaptation at substantially lower cost
than reported MILP-based synthesis times~\cite{shah2023taccl,liu2024teccl}.

\begingroup
\makeatletter
\renewcommand\section{\@startsection{section}{1}{\z@}%
  {-3.25ex plus -1ex minus -.2ex}%
  {1.5ex plus .2ex}%
  {\reset@font\large\bf}}
\makeatother
\section{Conclusion}
In this paper, we present \mydesign{}, an AlltoAllv framework for PCIe GPU systems that
separates topology-dependent channel analysis from demand-dependent scheduling. Its offline analyzer compresses
feasible channels into reusable families, while its lightweight online
scheduler decomposes each invocation's demand into channel activations without
repeating topology search.
Across four PCIe platforms and deployments of up to 256 GPUs, \mydesign{}
outperforms existing systems and reduces Qwen3 prefill and Wan2.1 generation
latency by up to 27.2\% and 6.1\%.

\endgroup

\bibliographystyle{plain}
\bibliography{refs}

\end{document}